\documentclass{aa}
\usepackage{graphics}

\begin{document}

\title{ Interstellar H$_2$ in M 33 detected with {\sc Fuse}\thanks{Based on observations made with the NASA-CNES-CSA Far Ultraviolet 
Spectroscopic Explorer, available in the public archive. FUSE is operated for NASA by the Johns Hopkins University under NASA contract NAS5-32985.}}

\author{H. Bluhm \inst{1}, K.S. de Boer \inst{1}, O. Marggraf \inst{1}, P. Richter \inst{2}, and B.P. Wakker \inst{3}}

\institute{Sternwarte, Universit\"at Bonn, Auf dem H\"ugel 71, 
53121 Bonn, Germany
 \and
 Osservatorio Astrofisico di Arcetri, Largo E. Fermi 5, 50125 Firenze, Italy
 \and
 University of Wisconsin--Madison, 475 N. Charter Street, Madison, WI 53706, USA}

\date{Received <date> / Accepted <date>}

\offprints{
\email{hbluhm@astro.uni-bonn.de}
}

\authorrunning{H. Bluhm et al.}
\titlerunning{Interstellar H$_2$ in M 33}

\abstract{ {\sc Fuse} spectra of the four brightest \ion{H}{ii} regions in \object{M~33} show absorption by interstellar gas in the Galaxy and in \object{M~33}. On three lines of sight molecular hydrogen in \object{M~33} is detected. This is the first measurement of diffuse H$_2$ in absorption in a Local Group galaxy other than the Magellanic Clouds.
 A quantitative analysis is difficult because of the low signal to noise ratio
 and the systematic effects produced by having multiple objects in the {\sc Fuse} aperture.
 We use the \object{M~33} {\sc Fuse} data to demonstrate in a more general manner the complexity of interpreting interstellar absorption line spectra towards multi-object background sources. 
 We derive H$_2$ column densities of $\approx10^{16}$ to $10^{17}$~cm$^{-2}$ along 3 sight lines (\object{NGC~588}, \object{NGC~592}, \object{NGC~595}).
 Because of the systematic effects, these values most likely represent upper limits and the non-detection of H$_2$ towards \object{NGC~604} does not exclude the existence of significant amounts of molecular gas along this sight line.
  \keywords{Galaxies: individual: M 33 -- ISM: abundances -- ISM: molecules  -- Ultraviolet: ISM}
}

\maketitle

\section{Introduction}

 Observations with the Copernicus satellite have shown that the ISM of the Milky Way contains large amounts of molecular hydrogen.
 The mass in H$_2$ may even dominate that in atomic form 
(see Savage et al. 1977). 
 Since 1996 other galaxies also have become accessible for investigation of the content of H$_2$, using {\sc Orfeus} and {\sc Fuse} far ultraviolet spectra. 
 Both the Large and the Small Magellanic Cloud appear to have substantial amounts of H$_2$ as found from absorption line studies (de Boer et al. \cite{deboer}; Richter et al. \cite{richter98}; Richter \cite{richter00}, Tumlinson et al. \cite{tumlinson}). 

 {\sc Fuse} has sufficient sensitivity to record spectra of bright sources in other nearby galaxies such as \object{M~31} and \object{M~33}. 
 It is of importance to extend our knowledge about the content in molecular gas to galaxies other than the Milky Way and the LMC and SMC. 
 These galaxies have different histories, different stellar contents and different abundances. 
 An intercomparison of these galaxies may lead to clues about the interplay of molecular cloud and star formation with chemical content and radiation field. 

 M\,33 is the smallest spiral galaxy of the Local Group at a distance of $\sim800$ kpc. 
It has a ragged optical appearance and bright H\,{\sc ii} regions, some of them  catalogued as NGC objects. 
The galaxy is inclined at 55$^{\circ}$ and shows signs of bending as 
derived from 21 cm radio synthesis observations (Rogstad et al. \cite{rogstad}). 
 In the direction of two of the bright H\,{\sc ii} regions 
CO has been found in emission (Wilson \& Scoville \cite{wilsco}). 
 The metallicity of M\,33 is subsolar with a radial abundance gradient which differs for different elements.
 For example, neon abundances in the observed \ion{H}{ii} regions vary between 0.2 and 0.3 dex subsolar (see  Willner \& Nelson-Patel \cite{willner}).
 Vibrationally excited H$_2$ has been found in emission by Israel et al.\ (1990) towards NGC~604, most likely from background gas.

 In this paper we investigate {\sc Fuse} spectra of four bright objects in M\,33. 
 Each of these consist of a number of young and thus UV-bright stars surrounded by an \ion{H}{ii} region. 
 In 3 of the 4 spectra H$_2$ is detected in absorption in \object{M~33}.

 In Sect. 2 we give some information about the instrument, the targets, and the data reduction, in Sect. 3 the difficulties with the interpretation of interstellar absorption spectra towards multi-object background sources are described. The measurements are presented in Sect. 4, followed by a discussion of the results in Sect. 5.  

\begin{figure*}
\resizebox{\hsize}{!}{\includegraphics{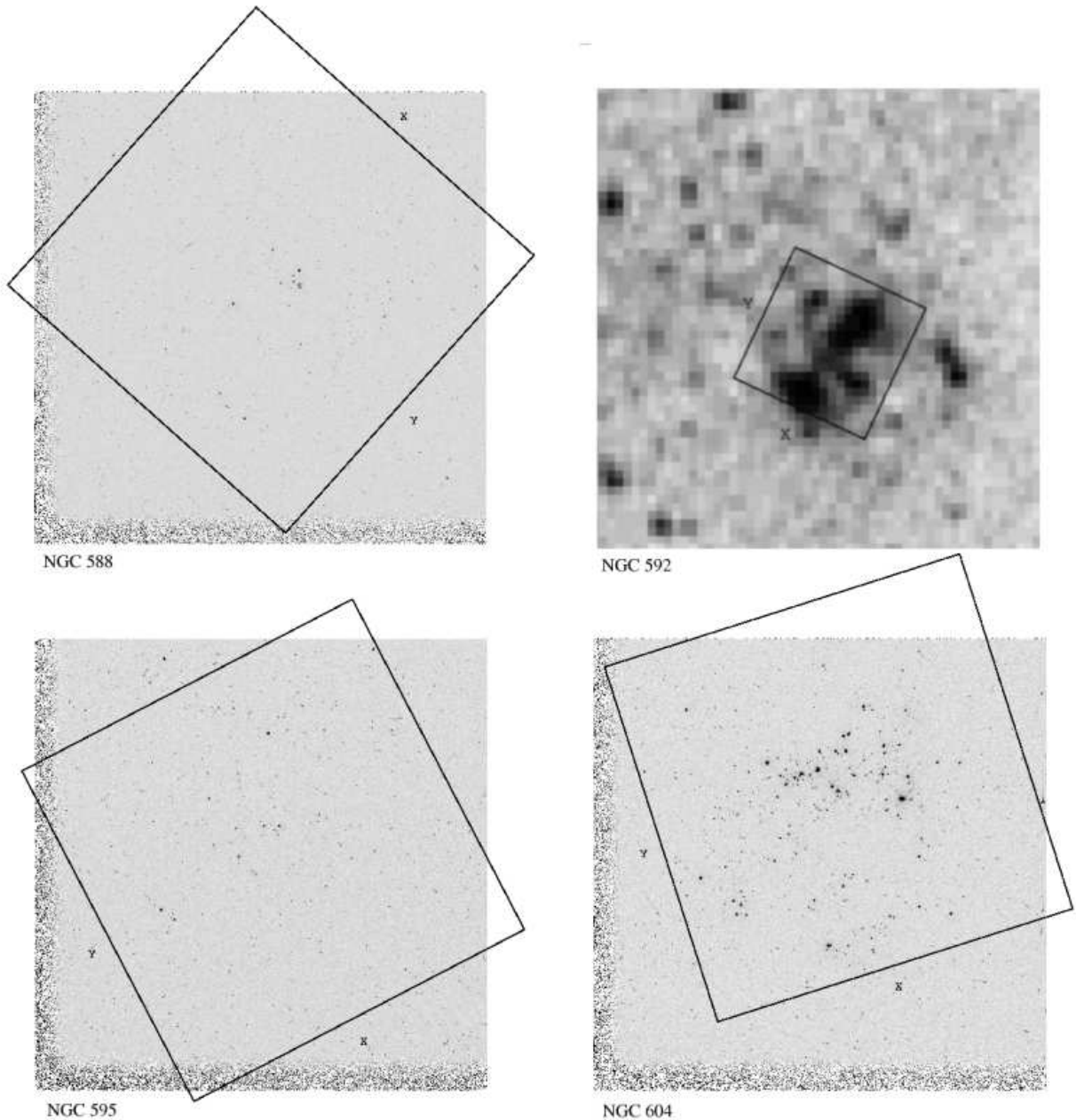}}
\hfill
\caption[]{The {\sc Fuse} $30 \times 30$ arcsec LWRS aperture overlaid on the four \ion{H}{ii} regions / star clusters. For \object{NGC~588}, \object{NGC~595}, and \object{NGC~604} HST WFPC 2 images  (central wavelength 1730~{\AA}) are shown, for \object{NGC~592} only a DSS1 image was available. }
\label{ap3}
\end{figure*}

\begin{table*}
\caption[]{Data on {\sc Fuse} pointings on \object{M~33} \ion{H}{ii} regions. The position angle is the angle of the {\sc Fuse} y axis (east of north). $N_{20}$ is the number of stars having an UV flux (at 1730~\AA) of at least $20\%$ of the brightest source in the area covered by the {\sc Fuse} aperture.
 For \object{NGC 592} $N_{20}$ is unknown because no public HST image is available (see Fig. \ref{ap3}). }
\begin{tabular}{lllllll}
\hline\noalign{\smallskip}
Object & Observation ID & $\alpha_{2000.0}$ & $\delta_{2000.0}$ & Pos. angle & Exp. time & $N_{20}$ \\
\hline\noalign{\smallskip}
NGC 588 & a0860404001 & 01\,32\,45.50 & +30\,38\,55.00 & $334.9^{\circ}$ & 2819~s & ~8 \\
        & a0860404002 & & & & 2401~s & \\
NGC 592 & a0860202001 & 01\,33\,12.30 & +30\,38\,49.00 & $334.9^{\circ}$ & 2164~s &  ~? \\
    & a0860202002  & & & & 1800~s & \\
NGC 595 & a0860303001 & 01\,33\,33.60 & +30\,41\,32.00 & $322.4^{\circ}$ & 3620~s & 12\\
NGC 604 & a0860101001 & 01\,34\,32.50 & +30\,47\,04.00 & $323.1^{\circ}$ & 3633~s & ~9\\    
    &  a0860101002 & & & & 3520~s & \\
\noalign{\smallskip}
\hline
\end{tabular}
\label{m33pointings}
\end{table*}

\section{Data}

\begin{figure*}
\resizebox{\hsize}{!}{\includegraphics{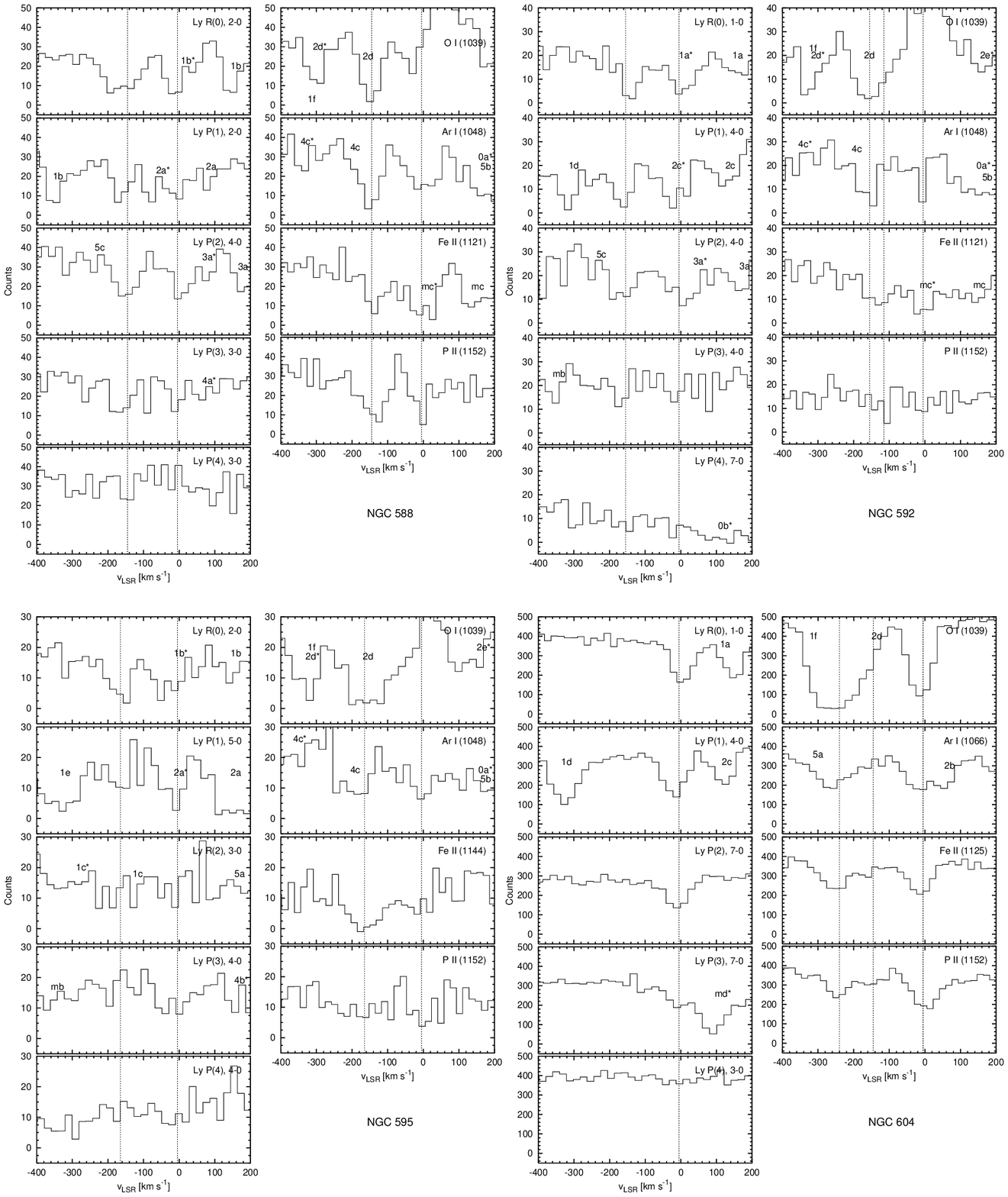}}
\hfill
\caption[]{For each line of sight a sample of five H$_2$ and four metal lines is shown. Because of peculiarities in the spectra (sight line structure, noise features, structures in the background source spectrum) the choice of transitions is not the same for all four sight lines. The dotted lines mark the  radial velocities of Galactic and M~33 gas.
 Identified other transitions within the range of the plots are labeled in the spectra as follows: H$_2$ Lyman: {\bf 0a}: R(0),4-0; {\bf 0b}: R(0),6-0; {\bf 1a}: R(1),1-0; {\bf 1b}: R(1),2-0; {\bf 1c} P(1), 3-0; {\bf 1d}: R(1),4-0; {\bf 1e}: R(1),5-0; {\bf 2a}: R(2),2-0; {\bf 2b}: P(2),3-0; {\bf 2c}: R(2),4-0; {\bf 2d}: R(2),5-0; {\bf 2e}: P(2),5-0; {\bf 3a}: R(3),4-0; {\bf 3b}: P(3),6-0;
{\bf 4a}: R(4),3-0; {\bf 4b}: R(4),4-0; {\bf 4c}; P(4),5-0; {\bf 5a}: P(5),4-0; {\bf 5b}: R(5),5-0; {\bf 5c}: P(5)(5-0). Metals: {\bf ma}: \ion{Ar}{i} (1048~\AA); {\bf mb}: \ion{Fe}{ii} (1055~\AA); {\bf mc}: \ion{Fe}{iii} (1122~\AA); {\bf md}: \ion{Si}{ii} (1020~\AA). \object{M~33} components are marked with a star.    
}
\label{spectra}
\end{figure*}

\begin{figure*}
\resizebox{\hsize}{!}{\includegraphics{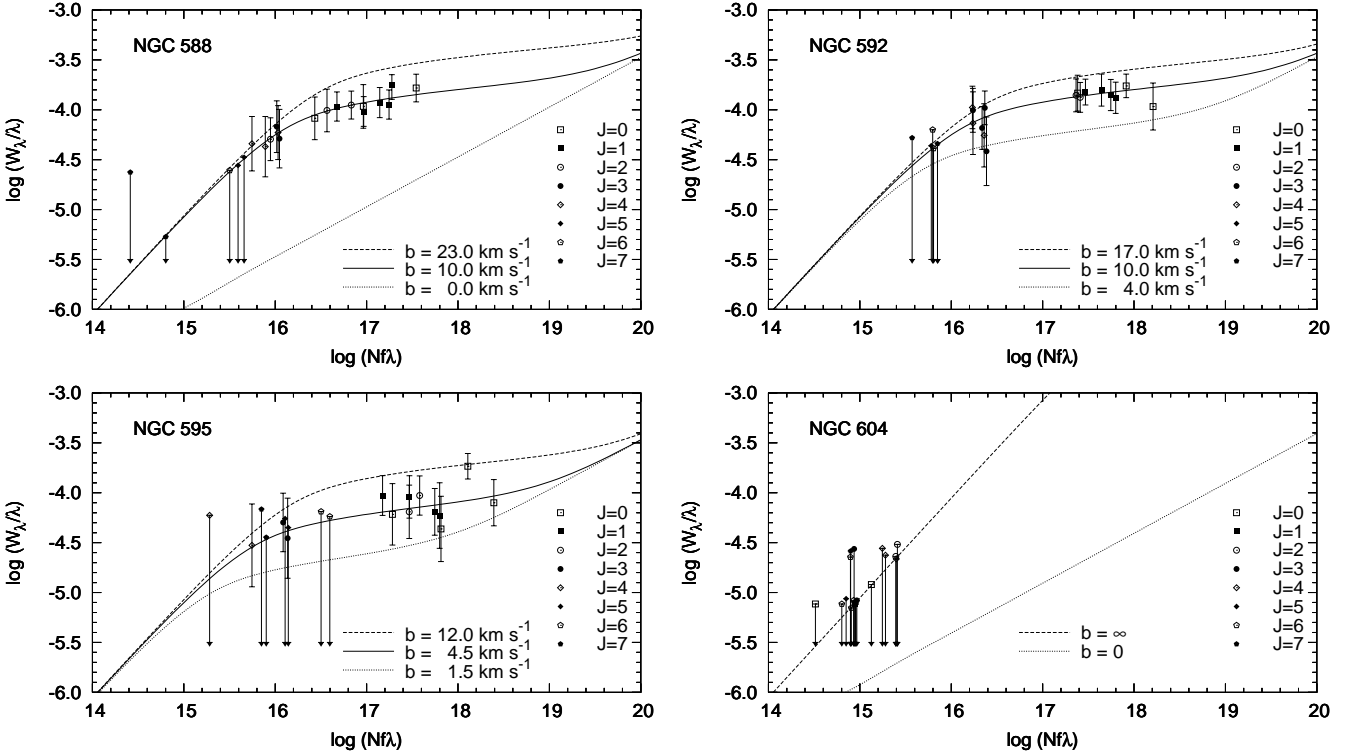}}
\hfill
\caption[]{
 Curves of growth for absorption by H$_2$ in \object{M~33}. The solid curves are for the best fit (or best guess) $b$-value, the position of the data points on the $\log(f\lambda)$ axis corresponds to the ``optimum" column densities given in Table \ref{h2colden}.
 The dashed and dotted curves represent the upper and lower limits of $b$-values consistent with the data.  
 They were used to determine the minimum and maximum column densities given in Table \ref{h2colden}.
}
\label{cog}
\end{figure*}

\begin{figure*}
\resizebox{\hsize}{!}{\includegraphics{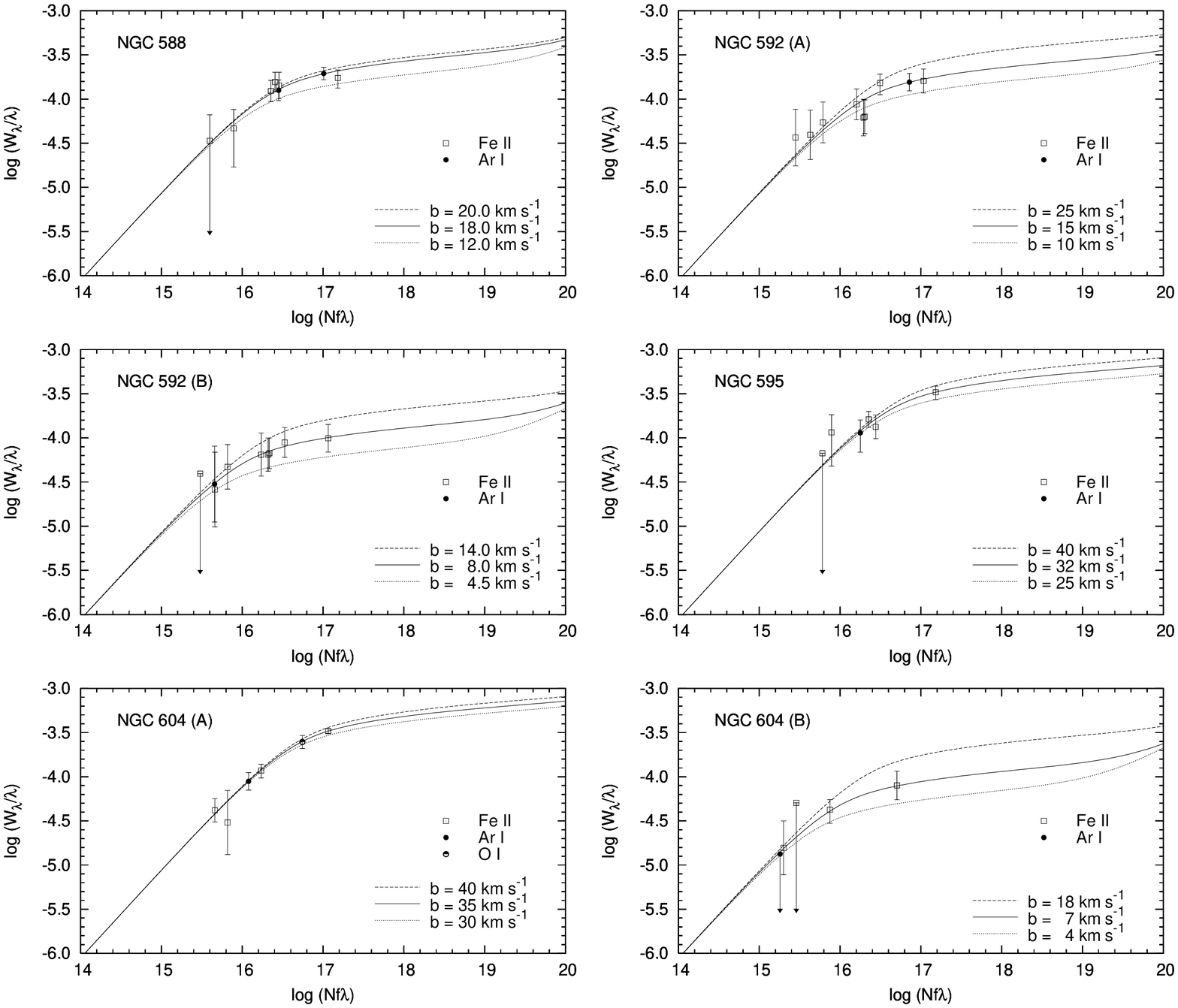}}
\hfill
\caption[]{Curves of growth for absorption by metals in \object{M~33}. Curves for the adopted $b$-value (solid lines), as well as for the upper and lower limits (dashed and dotted lines), are shown.
}
\label{cog.met}
\end{figure*}

\subsection{The instrument}
 {\sc Fuse} is equipped with four coaligned telescopes and Rowland spectrographs.
 The detectors are two microchannel plates.
 For a description of the instrument and its on-orbit performance see Moos et al. (\cite{moos}) and Sahnow et al. (\cite{sahnow}).
 Information about {\sc Fuse} data and their analysis can be found in the {\sc Fuse} Data Handbook\footnote{Available at http://fuse.pha.jhu.edu/analysis/dhbook.html}. 

 The typical velocity resolution of {\sc Fuse} spectra is about $20$~km\,s$^{-1}$.
 Because of variations in the alignment of the 8 different detector channels
the measurement of absolute radial velocities is problematic and requires a separate discussion (see Sect. 4.2). 
 
 The sensitivity threshold of {\sc Fuse} is set by the uncertainty in the background intensity level.
 It is below $(2 - 4)\times10^{-15}$~erg\,cm$^{-2}$\,s$^{-1}$\,{\AA}$^{-1}$.
 To allow a reasonable short integration time the source flux should be higher than about $1\times10^{-14}$~erg\,cm$^{-2}$\,s$^{-1}$\,{\AA}$^{-1}$.    

\subsection{The targets}

 The targets are the UV-bright cores of the \ion{H}{ii} regions \object{NGC 588}, \object{NGC 592}, \object{NGC 595}, and \object{NGC 604} in \object{M~33}. 
 For the {\sc Fuse} observations an entrance aperture with size of $30 \times 30$ square arcsec was used, being equivalent to $120\times 120$~pc$^2$ in \object{M~33}. 
 The precise locations of the apertures with their orientation and the exposure times are given in Table \ref{m33pointings}. 

 Inspection of imagery of the \object{M~33} \ion{H}{ii} regions shows that the {\sc Fuse} aperture has included numerous objects, of which only few were bright enough in the UV to have contributed to the background flux. 
 Nevertheless, the signal recorded is the sum of the flux of several stars. 
 This implies that the spectra provide a blend of spectra from different lines of sight with different line of sight gas (composition and velocity) structure, in particular within \object{M~33}. 

 In an attempt to account for this complexity, the imagery (Fig. \ref{m33pointings}) has been investigated to identify the brightest sources in the field. 
 We have included the scanty information from that in Table \ref{ap3}. 
 
 The complexity of the background sources complicates the interpretation of the interstellar absorption features as described in Sect. 3.
 Furthermore, the shape of the continua is difficult to assess in some cases.
 
 While the FUV fluxes of all 4 \ion{H}{ii} regions are well above the sensitivity of {\sc Fuse} ($\approx5\times10^{-14}$~erg\,cm$^{-2}$\,s$^{-1}$\,{\AA}$^{-1}$ for \object{NGC~588}, \object{NGC~592}, and \object{NGC~595}), the exposure times were too short to have more than $20$ to $50$ counts per $20$~km\,s$^{-1}$ resolution element. 
 An exception is the \object{NGC~604} spectrum where the flux ($\approx1\times10^{-12}$~erg\,cm$^{-2}$\,s$^{-1}$\,{\AA}$^{-1}$) provides roughly 400 counts per resolution element.
 Sample interstellar spectra are shown in Fig. \ref{spectra}, a few other spectra are given by Wakker et al. (\cite{wakker}). 
 Note that the spectra in Fig. \ref{spectra} are rebinned over 10 pixels.
 The resulting pixel size corresponds roughly to one resolution element.

\subsection{Data reduction}

 The raw spectra were calibrated with the Calfuse 1.8.7 reduction pipeline.
 For analysis the spectra were smoothed with a 5 pixel (NGC 604) or a 10 pixel boxcar filter (NGC 588, 592, 595) in order to handle the strong noise. 
 Equivalent widths were measured using trapezium or Gaussian fits.
 The errors were calculated taking into account the photon statistics and the uncertainty in the choice of the continuum.
 
 A calibration of the velocity zero point was made by comparison to radio data (see Sect. 4.2).

\section{General considerations for the interpretation}

\begin{figure}
\resizebox{\hsize}{!}{\includegraphics{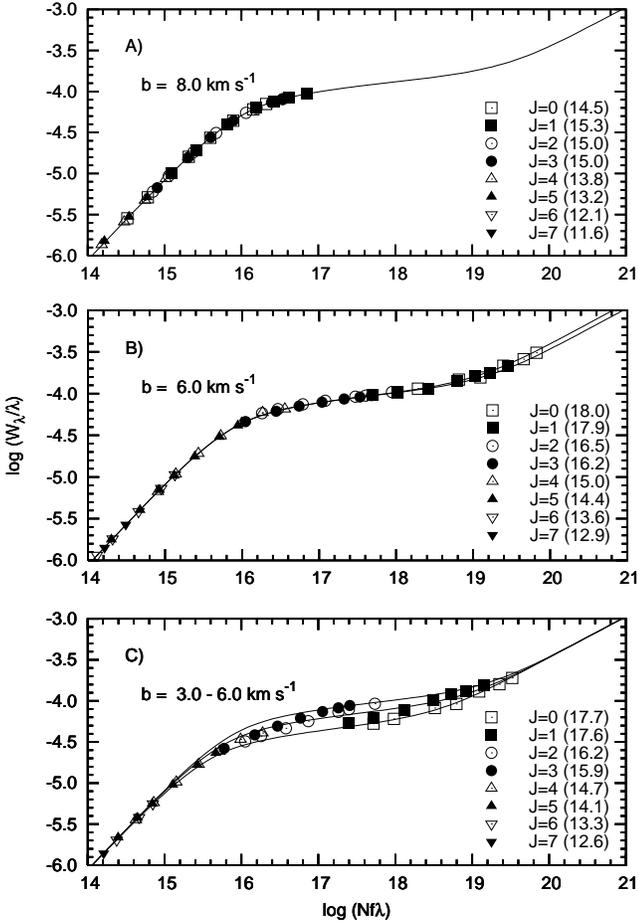}}
\hfill
\caption[]{The {\sc Fuse} spectra contain blends of the spectra of the brightest objects contained in the aperture. 
 The effects of such blends are explored assuming two equally bright background stars but one having a $2.6$~dex higher H$_2$ column density. 
 Panel {\bf A)} shows the curve of growth for the single sight-line spectrum 
with low H$_2$ column density, panel {\bf B)} the one for high H$_2$ column density. 
 Panel {\bf C)} shows the optical depth relation for the added spectrum. 
 In each panel the column densities of the rotational levels are indicated in
brackets, in {\bf C)} the mean values of the single sight line column densities are given. 
 It is obvious that in panel {\bf C)} a single cloud curve of growth interpretation is inappropriate.
}
\label{test}
\end{figure}

\subsection{Velocity structure}
  Due to the presence of several background targets within the {\sc Fuse} aperture the resulting spectrum is the sum of single spectra with different fluxes and absorption features.
 At the distance of \object{M~33} one arcsecond corresponds to a projected distance of about $4$~pc, so the {\sc Fuse} aperture covers an area of roughly $120\times 120$~pc$^2$.
 As Fig. \ref{ap3} shows, the stars which dominate the UV fluxes lie within smaller areas of about $50\times50$ to $70\times70$~pc$^2$. 
 On this scale significant fluctuations in velocity structure, $b$-values, and column densities of the \object{M~33} interstellar medium can be expected, especially for molecular gas that exists in confined regions.

 The velocity structure seen in such a sum spectrum thus reflects the combination of the possible presence of gas clouds with different velocities in front of the cluster, and/or possibly differing absorption velocities on the individual sight lines.

 For the Galactic foreground absorption the latter effect of widely separated background sources in \object{M~33} is probably of minor importance.
 While column density variations in the ISM are known to exist on small scales, velocity variations appear to be rather small (see e.g. Andrews et al. \cite{andrews}).
 
 Background sources separated in angle have yet another effect on the velocity structure of the spectrum.
 If the background sources are spread in the direction of dispersion, equal intrinsic absorption velocities and structures would be spread out to a maximum of about 100~km\,s$^{-1}$.
 As can be seen in Fig. \ref{spectra}, the H$_2$ and metal line absorption is wider than normally seen, but substantially less than the indicated maximum spread. 
 We conclude that the velocity smearing due to multiple background sources is less than $\approx50$~km\,s$^{-1}$. 

\subsection{Absorption strength}

 In order to judge in which way the blending of the light from the individual objects affects the interpretation of the spectra we will consider a simple but realistic case.

 For a single line of sight, several clouds and velocity components contribute to the interstellar absorption spectrum and the optical depth in an absorption profile is the sum of optical depths along the sight line. 

 In contrast, for a sum of spectra of different sight lines the measured equivalent widths of interstellar lines are mean values of the equivalent widths belonging to individual sight lines weighted by the flux of the background sources.

 We have constructed a simple case of blending to investigate in which way the data become less accessible for interpretation. 
 We consider two sight lines which contribute to one total spectrum by equal amounts. 
 The absorption structure of the H$_2$ lines on these lines of sight has been chosen to be different, one has a H$_2$ column density of $4.4\times10^{15}$~cm$^{-2}$, the other has $1.8\times10^{18}$~cm$^{-2}$. 
 Consequently, the lines in one spectrum are much weaker (if not even `absent'), and the light of that star blends some of the absorption present in the other spectrum.
 Clearly, the equivalent widths derived from the sum spectrum have been altered compared to those of the single spectra.

 Column densities determined from lines in the linear part of the curve of growth are the weighted mean of the column densities on the individual sight-lines.
 If lines in the damping part of the curve of growth are added, the measured column density is proportional to 
\begin{equation}
\frac{1}{\sum_{i}{g_i}}\cdot{\sum_{i}{(g_i{\cdot}W_i)})^2}.
\end{equation}
 This is less or equal to
\begin{equation}
\frac{1}{\sum_{i}{g_i}}\cdot{\sum_{i}{g_i}{\cdot}W_i^2},
\end{equation}
 the weighted mean of the individual column densities, where $W_i$ are the equivalent widths in the individual spectra and $g_i$ are the weights.
 In the logarithmic part of the curve, where the equivalent widths depend strongly on the $b$-value the measured column densities are again another kind of mean value.
 The situation is even more complicated if line profiles from different parts of the curve of growth are mixed.

\begin{table*}
\caption[]{ Column densities $\log N$ ($N$ in cm$^{-2}$) and velocities of \object{M~33} absorption components detected in the {\sc Fuse} spectra. Note that the H$_2$ column densities are ``best guess" values based on the ``optimum" $b$-values given in Table \ref{h2colden}.  }
\begin{tabular}{lllllll}
\hline\noalign{\smallskip}
 & NGC 588 & \multicolumn{2}{l}{NGC 592} & NGC 595 & \multicolumn{2}{l}{NGC 604} \\
\hline\noalign{\smallskip}
 &  & A & B &   & A & B \\
$v_{\rm LSR}$ [km\,s$^{-1}$] & $-140$ & $-150$ & $-110$ & $-160$ & $-235$ & $-140$ \\
\noalign{\smallskip}\hline\noalign{\smallskip}
\ion{Fe}{ii} & $15.10^{+0.30}_{-0.20}$ & $14.95^{+0.35}_{-0.15}$ & $15.00^{+0.60}_{-0.30}$ & $15.10^{+0.20}_{-0.10}$ & $15.00^{+0.10}_{-0.10}$ & $14.60^{+0.60}_{-0.30}$ \\
\ion{Ar}{i}  & $14.60^{+1.00}_{-0.25}$  & $14.45^{+2.25}_{-0.55}$ & $13.25^{+2.45}_{-0.50}$ & $14.40^{+0.25}_{-0.30}$ & $13.65^{+0.15}_{-0.10}$ & $<12.9$ \\
\ion{O}{i} & & & & & $16.20^{+0.30}_{-0.20}$ &  \\
H$_2$        & $16.6$ & $17.4$ & -- & $17.3$  & \multicolumn{2}{c}{$<15.0$} \\
\noalign{\smallskip}\hline\noalign{\smallskip}
\end{tabular}
\label{m33comp}
\end{table*}

  We illustrate the effects of the considered case in Fig.\,\ref{test}. 
  From this simple example, considering the resulting optical depth relation (see Fig.~\ref{test}~c), we conclude that column densities can be derived from the 
{\sc Fuse} \object{M~33} \ion{H}{ii} region spectra only in an approximate manner. 
 In particular, a single $b$-value cannot be used to interpret the absorption line strengths (as visible in Fig.~\ref{test}~c).

\subsection{Extinction}

 If the dust abundance in front of a cluster is spatially inhomogenous, a selection effect has to be considered for the interpretation of measured column densities.
 In general the dust abundance is correlated with the column density of the gaseous ISM. 
 Sight lines with higher gas column densities have a higher colour-excess, reducing the far UV flux in the spectra and thus lowering the weight in the sum of spectra. 
 Therefore, column densities measured in the composite spectrum of a cluster can be expected to be lower than the actual (un-weighted) average column densities towards the cluster.
 Because the abundance of H$_2$ is correlated with dust (see Savage et al. \cite{savage}, Richter \cite{richter00}, Tumlinson et al. \cite{tumlinson}), sight lines with high H$_2$ column densities are also expected to be suppressed in the composite spectrum.

\section{Results}

\subsection{Column densities}
\subsubsection{Molecular hydrogen}
 
 Despite the systematic errors described in Sect. 3 we use single cloud curves of growth for column density estimates because the present data do not allow a more detailed modelling. 
  For each rotational level the measured equivalent widths form a set of data points with unknown position on the $\log(Nf\lambda)$ axis.
 These sets are shifted along this axis in order to fit a single-cloud curve of growth.

 In the present case the range in $\log(f\lambda)$ covered by each set of data points is small and the uncertainties in the equivalent widths are large.
 Thus the ranges of possible $b$-values and column densities are substantial.   

 As mentioned in Sect. 3.2, in the case of a multiple background source, column densities derived from the damping part are systematically lower than the weighted mean of column densities for the individual sight lines.
 Thus the upper limits are not as rigid as they would be for a single sight line.

\subsubsection{Metals}

 The curve of growth method was also used to obtain \ion{Fe}{ii} and \ion{Ar}{i} column densities. 
 The results, especially for \ion{Fe}{ii}, can be assumed to be more reliable than those for H$_2$.
 Due to the usually higher $b$ - values in the warm neutral medium and the large range of $\log f\lambda$ - values of accessible \ion{Fe}{ii} lines some data points lie on the linear part of the curve of growth (see Fig. \ref{cog.met}).
 \ion{Fe}{ii} and \ion{Ar}{i} are the predominant ionization states of Fe and Ar in the neutral interstellar medium.
  The cosmic abundance ratio is $\log$(Fe/Ar)$\simeq1.1$.
 According to Sofia \& Jenkins (\cite{sofia}) argon is unlikely to be bound to dust in clouds that are thin enough to be observed in absorption. 
 In partially ionized clouds, significant amounts of argon can be ionized due to its large ionization cross section although its ionization energy lies above that of hydrogen. 
 Sofia \& Jenkins (\cite{sofia}) measured \ion{Ar}{i} depletions between $0.2$ and $0.6$ on Galactic sight lines. 
 Therefore, in general, the Fe/Ar ratio will yield an estimate for the iron depletion that is too low. 

 The degree of ionization of oxygen is coupled to that of hydrogen due to charge exchange reactions.
 Thus it would be desirable to know \ion{O}{i} column densities, which then could be converted into \ion{H}{i} column densities provided the metallicity is known.
 Unfortunately the $1039$~{\AA} line of \ion{O}{i} is affected by blending with H$_2$ lines and emission features and lies in the flat part of the curve of growth.
 We have included a plot of this line in Fig. \ref{spectra}, but do not analyse it.
 Only the spectrum of \object{NGC~604} has a sufficient count rate below 1000~{\AA} for a measurement of the \ion{O}{i} line at 976~{\AA}.

 Metal column densities are given in Table \ref{m33comp}.

\subsection{Velocities}

 Because of the possible apparent velocity shifts induced by multiple background sources, and the variations in the alignment of the optical channels of {\sc Fuse} we measured only velocity differences between Galactic and \object{M~33} absorption components.
 Noise and continuum uncertainties limit the precision to about $10$~km\,s$^{-1}$.

 Using data from the Leiden - Dwingeloo - Survey  (Hartmann \& Burton {\cite{hartburt}), the velocity of the Galactic \ion{H}{i} in the direction of \object{M~33} can be determined.
 We fit $v_{\rm LSR}=0$~km\,s$^{-1}$ and convert the measured {\sc Fuse} velocities to LSR velocities.

 We compared emission line data (CO, \ion{H}{i}, \ion{H}{ii} or other ions) from the literature with the {\sc Fuse} absorption line data to get hints on the location of the absorbing gas relative to the \ion{H}{ii} regions and the \object{M~33} disk. 
 Velocities of \ion{H}{i} 21~cm emission are taken from Rogstad et al. (\cite{rogstad}) who found two disk components, a ``main'' and a more patchy ``weak" one.
 These \ion{H}{i} ``main component" velocities are in agreement with velocities extracted from the work of Deul \& van der Hulst (\cite{deul}) based on observations with a 5 to 10 times higher angular resolution.
 Notably, no ``weak component" shows up in the Deul \& van der Hulst data at the positions of the \ion{H}{ii} regions.
 This may be due the patchiness of that component or to lower sensitivity of the higher resolution data.

\begin{table}[ht!]
\caption[]{M~33 H$_2$ column densities [cm$^{-2}$]. For none of the sight lines a definite $b$-value could be determined.  Thus column densities of the rotational levels for maximum,optimum,and minimum $b$-values [km\,s$^{-1}$] consistent with the data are given in the 3rd, 4th, and 5th column. 
 For \object{NGC~588} the damping limit was used to derive the upper limits of the column densities.
 Though the actual $b$-value is certainly $>0$~km\,s$^{-1}$, a vanishing Doppler parameter cannot be excluded by the data.   
 For NGC 604 upper limits were derived using the Doppler- and the damping-part of the curve of growth.
 }
\begin{tabular}{llrrr}
\hline\noalign{\smallskip}
 & $J$ & \multicolumn{3}{c}{$\log N$} \\
\hline\noalign{\smallskip}
                 &  & \multicolumn{1}{c}{$b_{\rm max}$} & \multicolumn{1}{c}{$b_{\rm opt}$}  & \multicolumn{1}{c}{$b_{\rm min}$}  \\
\object{NGC 588} &  & $23.0$~km\,s$^{-1}$ & $10.0$~km\,s$^{-1}$ & $0.0$~km\,s$^{-1}$ \\ 
\noalign{\smallskip}\hline\noalign{\smallskip}
                 & 0 &  $15.6$ &  {\bf 16.1} &  $18.3$ \\
                 & 1 &  $15.7$ &  {\bf 16.4} & $18.5$ \\
                 & 2 &  $15.2$ &  {\bf 15.6} & $18.1$ \\
                 & 3 &  $14.7$ &  {\bf 15.0} & $17.9$ \\
                 & 4 &  $14.6$ &  {\bf 14.9} & $17.8$ \\
                 & 5 &  $<14.5$ & {\bf $<$\,14.6} & $<16.8$ \\
                 & 6 &  $<14.5$ & {\bf $<$\,14.5} & $<16.7$ \\
                 & 7 &  $<13.7$ & {\bf $<$\,13.7} & $<15.2$ \\ 
\noalign{\smallskip}\hline\noalign{\smallskip}
\object{NGC 592} &   &  $17.0$~km\,s$^{-1}$ & $10.0$~km\,s$^{-1}$ & $4.0$~km\,s$^{-1}$ \\
\noalign{\smallskip}\hline\noalign{\smallskip}
                 & 0 &  $16.0$ & {\bf 17.1} & $18.4$ \\
                 & 1 &  $15.8$ & {\bf 16.9} & $18.4$ \\
                 & 2 &  $15.6$ & {\bf 16.4} & $18.2$ \\
                 & 3 &  $15.1$ & {\bf 15.4} & $17.2$ \\
                 & 4 &  $15.0$ & {\bf 15.2} & $17.1$ \\
                 & 5 &  $<14.9$ & {\bf $<$\,15.0} & $<15.5$ \\
                 & 6 &  $<14.6$ & {\bf $<$\,14.7} & $<15.2$ \\
                 & 7 &  $<14.7$ & {\bf $<$\,14.8} & $<15.3$ \\ 
\noalign{\smallskip}\hline\noalign{\smallskip}
\object{NGC 595} &   &  $12.0$~km\,s$^{-1}$ & $4.5$~km\,s$^{-1}$ & $1.5$~km\,s$^{-1}$ \\
\noalign{\smallskip}\hline\noalign{\smallskip}
                 & 0 &   $15.5$ &  {\bf 17.0} & $17.9$ \\
                 & 1 &   $15.7$ &  {\bf 16.9} & $18.1$ \\
                 & 2 &   $15.2$ &  {\bf 16.5} & $17.7$ \\
                 & 3 &   $14.8$ &  {\bf 15.1} & $17.1$ \\
                 & 4 &   $14.7$ &  {\bf 14.9} & $16.6$ \\
                 & 5 &   $<14.8$ & {\bf $<$\,15.1} & $<17.1$ \\
                 & 6 &   $<14.9$ &  {\bf $<$\,15.5} & $<17.3$ \\
                 & 7 &   $<14.6$ & {\bf $<$\,14.8} & $<16.7$ \\ 
\noalign{\smallskip}\hline\noalign{\smallskip}
\object{NGC 604} &   &  $\inf$ &   & $0$~km\,s$^{-1}$ \\
\noalign{\smallskip}\hline\noalign{\smallskip}
                 & 0 & $<13.7$ & & $<15.7$ \\
                 & 1 & $<14.3$ & & $<15.9$ \\
                 & 2 & $<14.4$ & & $<16.7$ \\
                 & 3 & $<14.0$ & & $<15.8$ \\
                 & 4 & $<14.4$ & & $<16.6$ \\
                 & 5 & $<14.0$ & & $<15.7$ \\
                 & 6 & $<13.8$ & & $<15.3$ \\
                 & 7 & $<13.8$ & & $<15.5$ \\ 
\noalign{\smallskip}\hline\noalign{\smallskip}
\end{tabular}
\label{h2colden}
\end{table}

\subsection{Comments on the sight lines}
\subsubsection{\object{NGC~588}}
 One \object{M~33} velocity component is detected in absorption by metals and H$_2$ in this direction.
 However further components could be hidden due to the limited resolution and the noise.
 The radial velocity of $-140$~km\,s$^{-1}$ is about the same as that of the weak \ion{H}{i} component seen in emission. 
 The velocity of the main \ion{H}{i} component ($-160$~km\,s$^{-1}$) is more similar to that of the emitting gas in the \ion{H}{ii} region ($-168$~km\,s$^{-1}$, see O'Dell \& Townsley \cite{odell}).
 The column densities of \ion{Fe}{ii} and \ion{Ar}{i} indicate some depletion of iron and thus the presence of dust.

\subsubsection{\object{NGC 592}}
 For this \ion{H}{ii} region no emission line results are available.
 The smoothed \ion{H}{i} velocity maps give about $-160$ and $-146$~km\,s$^{-1}$ for the ``main" and the ``weak" component, respectively.  
 The {\sc Fuse} absorption velocities of $-150$ and $-110$~km\,s$^{-1}$ are closer to the velocity of the ``weak component", but $-150$~km\,s$^{-1}$ is also consistent with the main \ion{H}{i} component velocity. 
 H$_2$ is found only in the $-150$~km\,s$^{-1}$ component.

\subsubsection{\object{NGC~595}}

 An absorption component at $\approx-160$~km\,s$^{-1}$  is detected in metal and H$_2$ lines.
 Absorption lines by higher rotational levels ($J=2,3,4$) are possibly shifted to more negative velocities by $10$ or $20$~km\,s$^{-1}$. 
 Despite the rather high ``best guess" value for the H$_2$ column density, the \ion{Fe}{ii} to \ion{Ar}{i} ratio points to mild depletion and relatively low dust abundance.
 CO emission observations by Wilson \& Scoville (\cite{wilsco}) revealed the existence of two molecular clouds ($v_{\rm LSR}=-186$~km\,s$^{-1}$ and $-189$~km\,s$^{-1}$) south of the cluster but within the region covered by the {\sc Fuse} aperture.
 However, clouds detectable in CO emission normally have so much dust that UV star light from behind should not get through.  
 So our data do not refer to such dense and opaque gas.

 The velocity of the main \ion{H}{i} emission component is $\approx-183$~km\,s$^{-1}$, that of the ``weak" \ion{H}{i} component is $\approx-155$~km\,s$^{-1}$.
 Possibly the $\approx-155$~km\,s$^{-1}$ component dominates in H$_2$ and metals while in the $\approx-183$~km\,s$^{-1}$ component the higher $J$-levels of H$_2$ contribute more (see, e.g., the similar case of HD~5980 in the SMC; Richter et al. \cite{richter98}).
 
\subsubsection{\object{NGC~604}}

 Two absorption components are found in metal lines: a strong one at $\approx-235$~km\,s$^{-1}$ and a weaker one at $\approx-140$~km\,s$^{-1}$.
 The velocity of the first one is consistent with the velocity of the main \ion{H}{i} emission component ($v\approx-236$~km\,s$^{-1}$), while the velocities of the weaker absorption component and the weak \ion{H}{i} emission component ($v\approx-206$~km\,s$^{-1}$) do not match.

 No absorption by H$_2$ at \object{M~33} velocities is present, although Wilson \& Scoville (\cite{wilsco}) detected CO emission south-east of the area covered by the {\sc Fuse} aperture at a velocity of $-242$~km\,s$^{-1}$.
 The LSR velocity of the \ion{H}{ii} region is $-246$~km\,s$^{-1}$ (O'Dell \& Townsley \cite{odell}), similar to that of CO shifted about $+10$~km\,s$^{-1}$ compared to the main \ion{H}{i} emission component.

 The comparatively high count rate in this spectrum allows the determination of a relatively rigid upper limit for the H$_2$ column density.
 The non-detection of H$_2$ in absorption is consistent with the measured \ion{Fe}{ii} and \ion{Ar}{i} column densities (see Table \ref{m33comp}). 
 In both velocity components the abundances do not indicate any depletion.
 For component A it was possible to measure the \ion{O}{i} line at 976~{\AA}.
 The \ion{O}{i} to \ion{Fe}{ii} ratio is consistent with a cosmic abundance ratio and no depletion of iron.

\section{Discussion}

\subsection{H$_2$ or no H$_2$}
 The results for \object{NGC~588}, \object{NGC~592}, and \object{NGC~595} are very similar: almost identical \ion{Fe}{ii} column densities, $\log$(\ion{Fe}{ii}/(\ion{Ar}{i}) between 0.3 and 0.7, indicating moderate iron depletion, and radial velocities consistent with the ``weak" \ion{H}{i} component and differing from the main disk component.

 The velocity difference between the H$_2$ detected in absorption and the emission from the \ion{H}{ii} regions and the molecular clouds in their vicinity indicates that the observed diffuse H$_2$ is probably not located near the \ion{H}{ii} regions.
 Towards \object{NGC~604} the velocity of the main absorption component is consistent with the velocity of the main \ion{H}{i} emission component. 
  While the iron column density is about the same as towards the other targets, no H$_2$ is found on this sight line.

 For \object{NGC~595} and \object{NGC~604} photometric results have been published.
 Wilson \& Matthews (\cite{wilma}) give $E(B-V)=0.35$ for \object{NGC~604} and $E(B-V)=0.25$ for \object{NGC~595}, based on Palomar 60 inch data.
 They state that the colour magnitude diagrams suggest the presence of ``significantly more differential reddening in \object{NGC~604} than in \object{NGC~595}".
 From  HST observations of \object{NGC~595} Malumuth et al. (\cite{malumuth}) find $E(B-V)=0.36\pm0.28$.
 We conclude that the average dust abundance towards both clusters is comparable, with a considerable scatter which is probably larger towards \object{NGC~604}. Thus the extinction-effect discussed in Sect. 3.3 is heavily influencing the apparent H$_2$ column densities in front of the clusters. 

 As the Galactic and Magellanic H$_2$ surveys show (Savage et al. \cite{savage}, Tumlinson et al. \cite{tumlinson}), on sight lines with an $E(B-V)$ of 0.2 or larger H$_2$ column densities between $10^{19}$ and $10^{21}$~cm$^{-2}$ are common.
 Though most of the $E(B-V)=0.36$ towards \object{NGC~595} is supposed to originate in \object{M~33} (Galactic $E(B-V)\approx0.06$, see Burstein \& Heiles \cite{buhei}), the estimated H$_2$ column density ($\approx2\times10^{17}$~cm$^{-2}$) is rather low.
 There is no sign of lines from $J=0$ or $J=1$ lying in the damping part of the curve of growth as it would be expected for high column densities.
 A rough estimate of the $E(B-V)$ of the FUV brightest stars in \object{NGC~595} based on the results published by Malumuth et al. (\cite{malumuth}) yields a value of about 0.1.       
 For a colour excess like this an H$_2$ column density of $\approx10^{17}$~cm$^{-2}$ is realistic.

 The photometric data for \object{NGC~604} are not that detailed as for \object{NGC~595} but it is likely that the extinction selection effect is even more significant. 
 If the scatter in extinction is indeed larger towards \object{NGC~604}, it is probable that there are some almost unreddened stars which outshine the rest of the cluster in the far UV.
 Independent of the actual average column density of H$_2$ in front of the cluster, the few sight lines with negligible amounts of dust or H$_2$ will dominate the UV spectrum and thus lead to a non-detection of H$_2$.
 
 Our results make clear that it is difficult to get precise quantitative information from absorption spectroscopy of extended or multiple background sources using a large spectrograph aperture.
 
\subsection{The bottom line}  
 For the first time diffuse molecular hydrogen has been detected in absorption in a local group galaxy farther away than the Magellanic Clouds.  
 Since star formation is going on in \object{M~33} and molecular clouds had been detected in emission by CO and excited H$_2$, the presence of diffuse H$_2$ is not unexpected.
 The estimated column densities are similar to those found for Milky Way and LMC along low-extinction sight lines.

 More reliable data of other sight lines are desirable to investigate the physical conditions in the diffuse interstellar medium of \object{M~33}, especially parameters like radiation field and H$_2$ formation rate.
 So far, 10 bright OB stars in \object{M~33} have been observed with {\sc Fuse}.
 Maybe in at least some of these lines of sight fewer stars lie in the field of view than in the data analysed here.
 Also, future work with the smaller MDRS or HIRS apertures of {\sc Fuse} may alleviate the extinction-selection problem encountered with the current data.        

\acknowledgements
H.B. is supported by the GK {\it The Magellanic Clouds and other dwarf galaxies} (DFG GRK 118).
O.M. was supported by grant Bo 779/24 from the Deutsche Forschungsgemeinschaft (DFG),  P.R. is supported by the Deutsche Forschungsgemeinschaft, B.P.W. was partially supported by NASA grant NAG5-9024.

\end{document}